\documentclass[pre,reprint,twocolumn,amsmath,10pt,amssymb,aps,superscriptaddress,showkeys,showpacs]{revtex4-2}\input epsf
\usepackage{bm}
\usepackage{soul}
\usepackage{graphicx}
\usepackage{makeidx}
\usepackage{chemarr}
\usepackage{multirow}
\usepackage{color}
\usepackage[normalem]{ulem}
\usepackage{txfonts}
\definecolor{darkgreen}{rgb}{0,0.6,0}
 \definecolor{orange}{rgb}{0.99,0.257,0}
\newcommand{\be}{\begin{equation}}
\newcommand{\ee}{\end{equation}}
\newcommand{\ba}{\begin{eqnarray}}
\newcommand{\ea}{\end{eqnarray}}

\def\ie{{\it i.e}~}

\def\epsilon{\varepsilon}

\def\ie{i.e.}

\def\beqr{\begin{eqnarray}}
\def\eqnr{\end{eqnarray}}
\def\beq{\begin{equation}}
\def\bc{\begin{center}}
\def\ec{\end{center}}
\def\eqn{\end{equation}}

\usepackage{amsmath}

\usepackage{graphicx}
\usepackage{dcolumn}
\usepackage{bm}

\usepackage{hyperref}

\hypersetup{colorlinks,linkcolor={blue},citecolor={blue},urlcolor={blue}}

\begin{document}

\preprint{APS/123-QED}
\title{Optimizing search processes in systems with state toggling: exact condition delimiting the efficacy of stochastic resetting strategy}


\author{Hillol Kumar Barman}
\email{hillol@iitb.ac.in}
\affiliation{Department of Physics, Indian Institute of Technology Bombay, Powai, Mumbai 400076, India}

\author{Amitabha Nandi}
\email{amitabha@phy.iitb.ac.in}
\affiliation{Department of Physics, Indian Institute of Technology Bombay, Powai, Mumbai 400076, India}

\author{Dibyendu Das}%
\email{dibyendudas@iitb.ac.in}
\affiliation{Department of Physics, Indian Institute of Technology Bombay, Powai, Mumbai 400076, India}


\begin{abstract}
Will the strategy of {\it resetting} help a stochastic process to reach its target efficiently, with its environment continually toggling between a strongly favourable and an unfavourable (or weakly favourable) state? A diffusive run-and-tumble motion, transport of molecular motors on or off a filament, and motion under flashing optical traps are special examples of such state toggling. For any general process with toggling under Poisson reset, we derive a mathematical condition for continuous transitions where the advantage rendered by resetting vanishes. For the case of diffusive motion with linear potentials of unequal strength, we present exact solutions which reveal that there is quite generically a \emph{re-entrance} of the advantage of resetting as a function of the strength of the strongly favourable potential. This result is shown to be valid for quadratic potential traps by using the general condition of transition.


\end{abstract}

\maketitle

\section{Introduction}
Systems with state toggling are abundant in nature. Typical examples are 
Markov processes toggling between two dynamical states
with Poisson switching times. Some relevant examples are bacterial locomotion switching between a run and a tumble state \cite{Howard_Berg}, dynamic microtubules switching between a growth and a shrinkage state \cite{Dogterom_1993}, processive molecular motors along microtubules toggling between biased and unbiased diffusive states \cite{Julicher_1997}, and DNA-promoter toggling between high and low transcriptionally active states \cite{raser_sc, raj_pcb}. Such processes may 
involve a target search -- reaching a spatial boundary, or crossing a threshold level, for the first time -- well-known as the {\it first-passage problem}  \cite{redner2001guide, Majumdar_2013review, orsanga, Roldan_2016, nayak, rijal2020, rijal2022}.

There is a wide interest in optimizing the first-passage times. Over a decade, {\it stochastic resetting} has been studied as an useful optimization strategy to expedite first passage \cite{EvansMajumdar_2011PRL,EvansMajumdar_2011IOP,Evans_2013,Durang_2014,Evans_2014,Christou_2015,Majumdar_Sabhapandit_2015,Meylahn_2015,Nagar_2016,Pal_2016,Bhat_2016,Reuveni2016,Pal_2017,Skilev_2017,Roldan_2017,Chchkin_2018,Pal_2019,Ali_2022,pal_2024}; in particular in chemical reactions \cite{Reuveni_2014,Reuveni_2015,Robin_2018}, biological processes \cite{Roldan_2016,Bressloff_2020_multiple_targets,Schumm_2021}, computational searches \cite{Montanari}, and magnetic phase transitions \cite{Magoni_2020}. The strategy involves repeatedly resetting and restarting the search process from the initial state, at random intervals of time, thus preventing the process from drifting too far away from the target. For exponentially distributed reset times, it was showed that there is an \emph{optimal reset rate} (ORR) at which mean first passage time (MFPT) is minimised \cite{EvansMajumdar_2011PRL}. 
The strategy has also been studied for non-instantaneous \cite{Bodrova_2020,pal_2024}, and non-Poissonian \cite{Bhat_2016,Pal_2017, Nagar_2016, Pal_2016,Skilev_2017} resets, and recently investigated in experiments as well \cite{Friedman_2020,Besga_2020,Somnath_PRXLife}.

Is resetting strategy always helpful during target search or can it be harmful sometimes?  In fact this interesting question has been studied in the presence of variety of attractive as well as barrier potentials. Across boundaries in parameter space, the ORR characterizing the optimal advantage of resetting has been shown to vanish continuously or discontinuously \cite{Christou_2015,Saeed_2019,Saeed_2020,Saeed_2022,Saeed_2023, Ray_2019,Pal_2019_Landau_like,Ray_2020,RKSingh_2020,Pal_2020}. Landau like theories have been developed for such transitions \cite{Saeed_2019,Pal_2019_Landau_like,Saeed_2022}. 
An important result in this context was derived to locate the  continuous ORR vanishing transitions in parameter space \cite{Saeed_2019, Pal_2019_Landau_like, Saeed_2022}, namely,
\begin{equation}
\sigma^2=\langle T \rangle^2,\label{Transition1}
\end{equation}
where variance $\sigma^2$ and MFPT $\langle{T}\rangle $ are in the {\it absence of resetting}. This condition is rather useful, as often the problem with resetting may be difficult to solve, compared to the problem without resetting. 

The same question of \emph{efficacy of resetting during target search} may be asked for systems with \emph{two-state toggling}. In this paper, we address this question for Poisson resets and derive a  general condition analogous to Eq.~\ref{Transition1}, applicable to such systems.

Previous studies have looked at transport properties in model systems with state toggling \cite{Tailleur_2008,Elgeti_2015,Angelani_2015,Angelani_2017,Slowman_2017,angelani2014first,Rupprecht_2016,Dhar_2019,Mercado-Vásquez_2020,Mori_2020,Bressloff_2020,Santra_2020,Mercado_2021,PSPal_2024,Malakar_2018,Evans_2018_RTP,Doussal_2021,Santra_2021,Roberts_2023,santra_2024}, and some specifically into first passage problems \cite{angelani2014first,Rupprecht_2016,Malakar_2018,Dhar_2019,Mercado-Vásquez_2020,Mori_2020,Bressloff_2020,Santra_2020,Mercado_2021,PSPal_2024}. The first passage of a run-and-tumble particle was studied without resetting \cite{Malakar_2018}, and also in the presence of resetting \cite{PSPal_2024,Evans_2018_RTP}, with and without diffusion. Interestingly, a feature relevant to run-and-tumble dynamics was not considered in these earlier works, namely the effect of a chemical gradient which makes  the \emph{toggling rates unequal} with preference towards the food source (target). We show below, that unequal rates reveal a part of parameter space with interesting ORR vanishing transitions.

In this paper, (i) we derive a general condition to locate continuous ORR vanishing transitions in first passage problems of any stochastic process with state toggling. (ii) We solve the MFPT for the general diffusive problem toggling between two arbitrary `linear' potentials at unequal rates, under resetting, and also demonstrate the applicability of the general condition analytically. The results reveal a generic feature of \emph{re-entrance} in problems with state toggling -- the advantage of resetting vanishes and then reappears as a function of strength of the attractive bias. {(iii) We discuss the relevance of the result in a biophysical problem involving  molecular motor assisted cargo transport.} (iv) The re-entrance phenomenon is also demonstrated for flashing nonlinear quadratic traps. Such re-entrance has been reported in systems without toggling under special conditions -- with two absorbing boundaries, or a barrier separating two valleys \cite{Saeed_2022, Pal_2019_Landau_like}; its generic appearance in systems with two-state toggling thus assumes importance.\\

\section{Two-state toggling process: derivation of the general condition for continuous transition}

\begin{figure}[hbt!]
\includegraphics[scale=0.38]{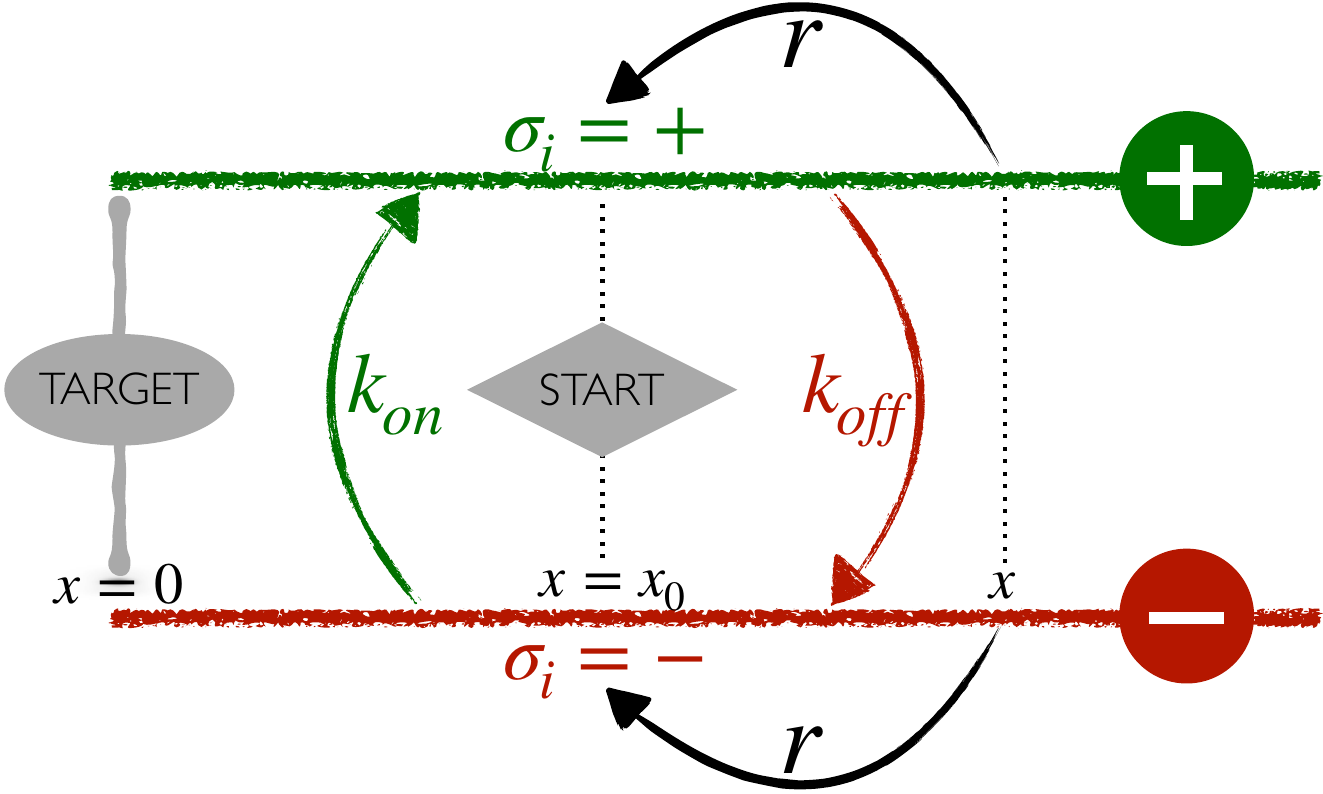}
\caption{
Schematic diagram depicting the 
target ($x=0$) search of a stochastic process with coordinate $x$, toggling between two states ``$+$" and ``$-$" at dissimilar rates $k_{off}$ and $k_{on}$, in the presence of resetting at rate $r$, to its initial coordinate $x_0$. 
}
\label{fig1}
\end{figure}

We consider a general two-state toggling process schematically depicted in Fig.\ref{fig1}. The dynamic process with a generalized stochastic coordinate $x(t)$ (which for example can be a position, or a chemical reaction coordinate) searches for a target at $x=0$, intermittently toggling between two states $\sigma(t) = +$ and $\sigma(t) = -$ at rates $k_{off}$ and $k_{on}$. The process is occasionally reset back to its initial coordinate $x = x_0$ at rate $r$, without changing the instantaneous state $\sigma(t)$. 



Since we are interested in first passage statistics, it is natural to start with backward stochastic equations for survival probability \cite{gardiner} $Q_r^{\sigma_i}(x_0,t)$ i.e. the probability of a particle initially at $x_0$ in state $\sigma_i$, having survived (not reached $x=0$) up to time $t$, under resetting at rate $r$. This is done in Sec \ref{secIII}. In this section, we \sout{wish to} start with relevant renewal equations. {We note that in biophysical scenarios like bacterial run-and-tumble or motor transport, a positional reset (i.e. $x(t) \rightarrow x_0$) need not happen with a state reset (i.e. $\sigma(t) \rightarrow \sigma_i$) and thus a complete renewal does not happen on resetting.} The general mathematical possibility of simultaneous switch of the state $\sigma(t)$ with probability $\eta$ along with positional reset,  was considered in  \cite{Evans_2018_RTP}. Here we consider the limit $\eta = 0$ given the biophysical motivations mentioned above. {Since the process is not fully renewed upon resetting, it helps to develop the formalism  in terms of \emph{joint survival probability}   $Q_{r}^{\sigma_f\sigma_i}(t)$ \cite{Evans_2018_RTP} -- it denotes the joint probability of not reaching $x=0$ up to time $t$ just like $Q_r^{\sigma_i}(x_0,t)$ along with the requirement $\sigma(t) = \sigma_f$.} The subscript $r$ indicates the presence of resetting, and in the absence of resetting the subscript will be set to 0. A set of renewal equations for 
$Q_{r}^{\sigma_f\sigma_i}(t)$ can be written for this general two state toggling problem (see Fig. \ref{fig1}) as follows:

\begin{eqnarray}
    Q_r^{\sigma_f\sigma_i}(t)&=&e^{-rt}Q_0^{\sigma_f\sigma_i}(t)\nonumber\\
    & &+r\int_0^t d\tau e^{-r\tau}\Big[Q_0^{\sigma_f\sigma_f}(\tau) Q_r^{\sigma_f\sigma_i}(t-\tau)\nonumber \\& & +Q_0^{\sigma_f,-\sigma_f}(\tau)Q_r^{-\sigma_f\sigma_i}(t-\tau) \Big].\label{renewal1}
\end{eqnarray}
In Eq. (\ref{renewal1}), the first term arises for survival without reset up to time $t$. The remaining terms are for survival with reset, where the last reset occurred at time $(t-\tau)$ and the state being $\sigma_f$ (in the second term) or $-\sigma_f$ (in the third term) at that moment. Defining Laplace transforms $\Tilde{Q}_{r,0}^{\sigma_f\sigma_i}(s)=\int_0^\infty Q_{r,0}^{\sigma_f\sigma_i}(t) e^{-st}dt$, Eq. (\ref{renewal1}) leads to the following in the Laplace domain (see also \cite{Evans_2018_RTP}):
\begin{eqnarray}
    \Tilde{Q}_r^{\sigma_f\sigma_i}(s)&=&\Tilde{Q}_0^{\sigma_f\sigma_i} (r+s)+r\Big[\Tilde{Q}_0^{\sigma_f\sigma_f}(r+s)\Tilde{Q}_r^{\sigma_f\sigma_i}(s)\nonumber\\
    & &+\Tilde{Q}_0^{\sigma_f,-\sigma_f}(r+s)\Tilde{Q}_r^{-\sigma_f\sigma_i}(s)\Big]. \label{Lap_renewal}
\end{eqnarray}\\
Eq.~\ref{Lap_renewal} represents four coupled equations (Eq.~\ref{explicit_eqs})
and lead to the solutions for $\Tilde{Q}_r^{\sigma_f\sigma_i}(s)$ in the presence of resetting, in terms of $\Tilde{Q}_0^{\sigma_f\sigma_i}(r+s)$ in the absence of resetting (Eq.~\ref{Qr(s)}).

The survival probabilities with initial state ($\sigma_i,x_0$), are related to the joint probabilities  as $Q_r^{\sigma_i}(t)=\sum_{\sigma_f=\pm} Q_r^{\sigma_f\sigma_i}(t)$ and $Q_0^{\sigma_i}(t)=\sum_{\sigma_f=\pm} Q_0^{\sigma_f\sigma_i}(t)$, with and without resetting respectively. Consequently, the Laplace transforms $\Tilde{Q}_r^{\sigma_i}(s)=\sum_{\sigma_f=\pm} \Tilde{Q}_r^{\sigma_f\sigma_i}(s)$ and $\Tilde{Q}_0^{\sigma_i}(s)=\sum_{\sigma_f=\pm} \Tilde{Q}_0^{\sigma_f\sigma_i}(s)$. The MFPTs of our interest are obtained as $\langle T_r\rangle^{\sigma_i}=\Tilde{Q}_r^{\sigma_i}(s\rightarrow0)$
:
\begin{eqnarray}
   &&\langle T_r \rangle ^{+}=\nonumber\\
   &&\frac{\Tilde{Q}_0^{+}(r)+r[\Tilde{Q}_0^{+-}(r)\Tilde{Q}_0^{-+}(r)-\Tilde{Q}_0^{--}(r)\Tilde{Q}_0^{++}(r)]}{1-r[\Tilde{Q}_0^{++}(r)+\Tilde{Q}_0^{--}(r)]+r^2[\Tilde{Q}_0^{++}(r)\Tilde{Q}_0^{--}(r)-\Tilde{Q}_0^{+-}(r)\Tilde{Q}_0^{-+}(r)]}\ ,\nonumber\\\nonumber\\
   &&\langle T_r \rangle ^{-}=\nonumber\\
   &&\frac{\Tilde{Q}_0^{-}(r)+r[\Tilde{Q}_0^{+-}(r)\Tilde{Q}_0^{-+}(r)-\Tilde{Q}_0^{--}(r)\Tilde{Q}_0^{++}(r)]}{1-r[\Tilde{Q}_0^{++}(r)+\Tilde{Q}_0^{--}(r)]+r^2[\Tilde{Q}_0^{++}(r)\Tilde{Q}_0^{--}(r)-\Tilde{Q}_0^{+-}(r)\Tilde{Q}_0^{-+}(r)]}.\nonumber\\\label{Tasr}
\end{eqnarray}
 Note that the two MFPTs $\langle T_r\rangle^{+}$ and $\langle T_r\rangle^{-}$ are distinct functions of $r$ and therefore would have two distinct minima at two ORRs $r_*^+$ and $r_*^-$. As the ORRs vanish (\ie~$r_*^\pm\rightarrow 0$) continuously at a transition point, the coefficients of the first order terms in the Taylor expansion of $\langle T_r \rangle^\pm=a_0^\pm+ra_1^\pm+O(r^2)$ need to vanish, i.e.  $a_1^\pm=\frac{d\langle T_r\rangle^\pm}{d r}\Big |_{r_*^\pm\rightarrow 0} = 0$. This leads to the conditions 
(Eq.~\ref{Trcn}) for vanishing of ORR 
corresponding to $\langle T_r\rangle^{+}$ and $\langle T_r\rangle^{-}$, jointly written (with $\sigma_i=\pm$) as
\begin{eqnarray}
\label{ORR_cond}
    &&\big(\langle T \rangle ^{\sigma_i\sigma_i}\big)^2+ \langle T \rangle ^{-\sigma_i\sigma_i}\Big( \langle T \rangle ^{\sigma_i-\sigma_i}+ \langle T \rangle ^{-\sigma_i-\sigma_i}\nonumber\\
    &&+ \langle T \rangle ^{\sigma_i\sigma_i}\Big)-\frac{1}{2}\Big( \langle T^2 \rangle ^{\sigma_i\sigma_i}+ \langle T^2 \rangle ^{-\sigma_i\sigma_i}\Big)=0.\label{trn_condition}
\end{eqnarray}
Eq.~(\ref{ORR_cond}) is a general condition applicable to any stochastic process with toggling. It reduces to  the condition Eq.~(\ref{Transition1}) for systems without toggling, by setting the last term in Eq.~(\ref{renewal1}) to zero which leads to $\langle T \rangle ^{\sigma_i-\sigma_i} = \langle T \rangle ^{-\sigma_i-\sigma_i}=0$ in Eq.~\ref{trn_condition}, and finally dropping the $\sigma_i$ superscripts. Note that the quantities involved in the absence of resetting in Eq.(\ref{trn_condition}), namely $\langle T\rangle^{\sigma_f\sigma_i}=\Tilde{Q}_0^{\sigma_f\sigma_i}(r)\big|_{r\rightarrow 0} = \int_0^\infty dt Q_0^{\sigma_f\sigma_i}(t)$ 
and $\langle T^2\rangle^{\sigma_f\sigma_i}=-2\partial_r \Tilde{Q}_0^{\sigma_f\sigma_i}(r)\big|_{r\rightarrow 0} = 2\int_0^\infty tQ_0^{\sigma_f\sigma_i}(t)dt$ are not ordinary moments of the first passage time but are related to the joint probabilities. They may be evaluated either exactly using  $\Tilde{Q}_0^{\sigma_f\sigma_i}(s)$ or in simulations using  $Q_0^{\sigma_f\sigma_i}(t)$, as we show below.\\ 

We discuss below several examples in this work to demonstrate the applicability of the general condition Eq.~\ref{ORR_cond}.

\begin{figure*}[hbtp]
\includegraphics[scale=0.5]{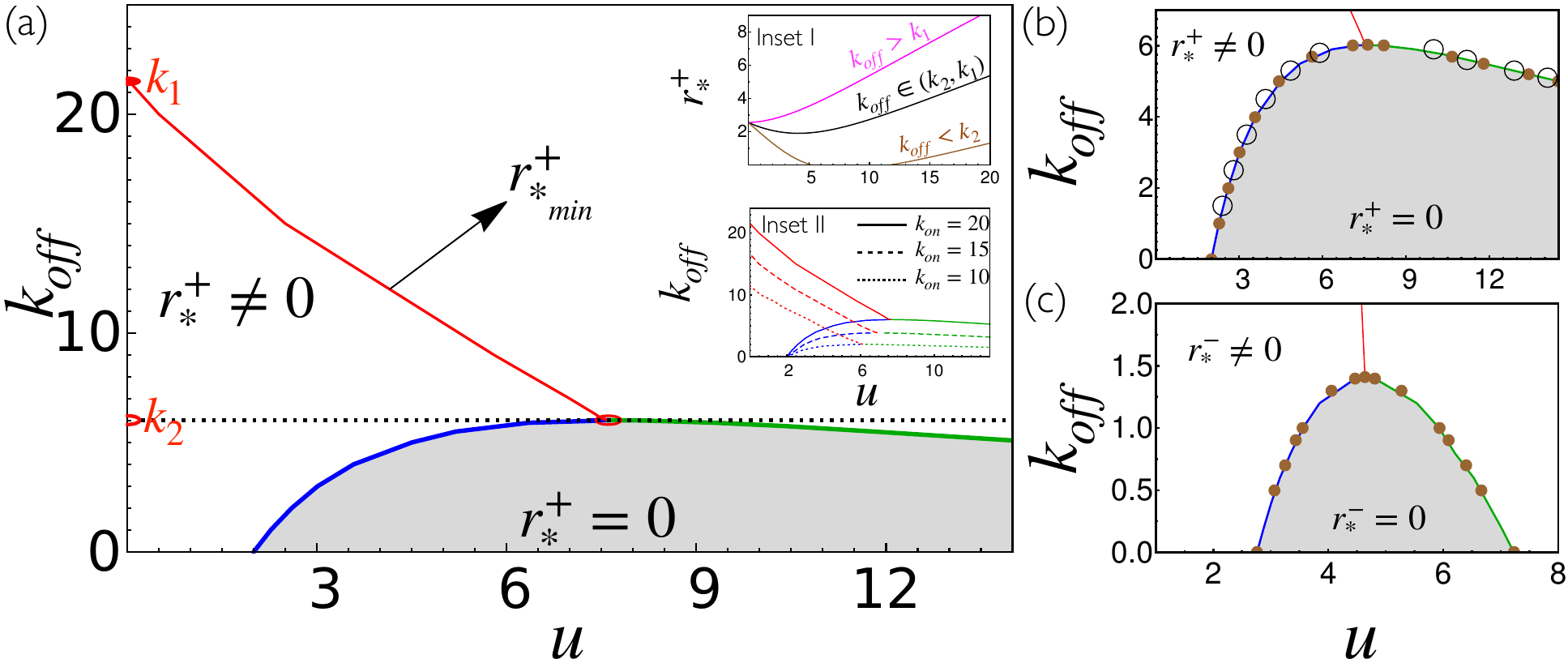}
\caption{(a) Phase diagram for $r_{*}^+$ corresponding to $\langle T_r\rangle^+$ in the $k_{off}-u$ plane for a run-and-tumble particle for $k_{on}=20$. The blue and the green lines representing the first and second transitions, determine the phase boundary between the regions $r_{*}^+=0$ (gray) and $r_{*}^{+}\neq0$ (white). The red line is the line of minimum resetting rate with its two end points at $k_{off}=k_1$ and $k_2$ (dotted line is a guide to the eye). 
Inset I shows ORR $r_*^+$ as a function of $u$ for different ranges of $k_{off}$ (with respect to $k_1=21.4$ and $k_2=6.1$). 
Inset II: Phase-diagram for $r_{*}^+$ similar to (a) for different $k_{on}$ values. (b) Zoomed-in view of the phase boundary in (a). The phase boundary obtained using the mathematical condition {Eq.~(\ref{trn_condition})} and via simulations are shown by brown filled circles and black hollow circles respectively. (c) Phase-diagram similar as (b), for $r_{*}^-$ corresponding to $\langle T_r\rangle^-$ for $k_{on}=20$. Brown circles on the phase boundary are obtained using {Eq.~(\ref{trn_condition})}. 
}
\label{f2}
\end{figure*}

\section{Two linearly biased states}
\label{secIII}

 We begin by considering the general problem of a diffusing particle in the domain $x \in [0,\infty)$ with diffusion constant $D$ under Poisson resetting (with rate $r$) and 
alternately subjected to a potential $V_+(x)=v_+x$ (the ``+" state) and a potential $V_-(x)=-v_-x$ (the ``$-$" state), with absorbing boundary at $x=0$. The toggling rates between the states are $k_{on}$ and $k_{off}$ respectively. As we show below, this linear problem can be exactly solved for the MFPTs. Note $v_+>0$ represents an attractive bias towards the target and $v_- > 0$ represents a repulsive bias away from the target. 

The survival probabilities $Q_r^+(x,t)$ and $Q_r^-(x,t)$ of not reaching $x=0$ up to time $t$, starting with $\sigma_i=+$ and $\sigma_i=-$ respectively, satisfy the backward Chapman-Kolmogorov equations \cite{gardiner}
\beqr
\partial_t Q_r^+(x,t)&=& D\partial_x^2 Q_r^+(x,t)-v_+\partial_x Q_r^+(x,t)-rQ_r^+(x,t)\nonumber \\
&+&rQ_r^+(x_0,t)+k_{off}\left[Q_r^-(x,t)-Q_r^+(x,t)\right], \nonumber\\
\partial_t Q_r^-(x,t)&=&D  \partial_x^2 Q_r^-(x,t)+v_-\partial_x Q_r^-(x,t)-rQ_r^-(x,t) \nonumber \\
&+&rQ_r^-(x_0,t)+k_{on}\left[Q_r^+(x,t)-Q_r^-(x,t)\right]. \nonumber \\
&& \label{survival_general_0}
\eqnr
The first two terms on the right hand side of each of the equations above  account for diffusion and drifts, the third and fourth terms for resetting, and the remaining terms for state toggling. Limiting cases of the above general problem have been studied in the literature earlier, namely run-and-tumble 
particle with $D=0, v_+=v_-, k_{on}= k_{off}$ in \cite{Evans_2018_RTP}, and 
$v_+ = v_-, k_{on}= k_{off}, r=0$ in   \cite{Malakar_2018}. Note that the special case of flashing potential is given by $v_+ > 0$, and $v_-=0$.


The initial condition implies $Q_r^\pm(x,0)=1$ at any $x>0$, and the absorbing boundary condition is $Q_r^\pm(0,t)=0$. We first non-dimensionalize these equations by rescaling: $t D/x_0^2\rightarrow t$, ~ $x/x_0\rightarrow x$, ~ $x_0v_{\pm}/D\rightarrow u_\pm$, ~
$k_{off}x_0^2/D\rightarrow k_{off}$, ~ $k_{on}x_0^2/D\rightarrow k_{on}$, ~
$rx_0^2/D\rightarrow r$ 
(see appendix \ref{A2}). Thus the resetting point $x=x_0$ now goes to $x=1$. Using Laplace transforms, a pair of inhomogeneous coupled equations for  $\Tilde{Q}_r^\pm(x,s)=\int_0^\infty Q_r^\pm(x,t) e^{-st}dt$ are obtained, which can be made homogeneous by shifting the functions by suitable constants. By making ansatz of exponential functions in space, and using the absorbing boundary conditions 
(appendix \ref{A2}), we solve for $\Tilde{Q}_r^\pm(1,s)$. The exact MFPTs $\langle T_r\rangle^\pm=\Tilde{Q}_r^\pm(1,s\rightarrow 0)$ are given in 
Eqs.~\ref{exact_t+}, \ref{exact_t-}-- these expressions are used to study ORR and its transitions below.

To study the validity of Eq.(\ref{trn_condition}) in this problem, we start with the equations for the joint probabilities $Q_0^{\sigma_f\sigma_i}(t)$, in the {\it absence} of resetting:
\beqr
\partial_t Q_0^{++}(x,t)&=&\partial_x^2 Q_0^{++}(x,t)-u_+\partial_x Q_0^{++}(x,t) \nonumber\\
&+&k_{off}\left[Q_0^{+-}(x,t)-Q_0^{++}(x,t) \right], \nonumber\\
\partial_t Q_0^{-+}(x,t)&=&\partial_x^2 Q_0^{-+}(x,t)-u_+\partial_x Q_0^{-+}(x,t)\nonumber\\
&+&k_{off}\left[Q_0^{--}(x,t)-Q_0^{-+}(x,t)\right], \nonumber\\
\partial_t Q_0^{+-}(x,t)&=&\partial_x^2 Q_0^{+-}(x,t)+u_-\partial_x Q_0^{+-}(x,t) \nonumber\\
&+&k_{on}\left[Q_0^{++}(x,t)-Q_0^{+-}(x,t)\right], \nonumber\\
\partial_t Q_0^{--}(x,t)&=&\partial_x^2 Q_0^{--}(x,t)+u_-\partial_x Q_0^{--}(x,t) \nonumber\\
&+&k_{on}\left[Q_0^{-+}(x,t)-Q_0^{--}(x,t)\right], \label{gen_wr}
\eqnr
with initial conditions $Q_0^{++}(x,0)=Q_0^{--}(x,0)=1$ and $Q_0^{-+}(x,0)=Q_0^{+-}(x,0)=0$ at any $x>0$, and boundary conditions $Q_0^{\pm\pm}(0,t)=0$. The Laplace transforms $\Tilde{Q}_0^{\sigma_f\sigma_i}(s)$ are solved exactly, leading to the relevant moments $\langle T^k\rangle^{\sigma_f\sigma_i}=(-1)^{k-1} k\dfrac{\partial^{k-1} \Tilde{Q}_0^{\sigma_f\sigma_i}(s)}{\partial s^{k-1}}\big|_{s\rightarrow 0}$ 
(see appendix~\ref{A3}) needed to verify Eq.(\ref{trn_condition}).
Before presenting the results for the general case of $u_+ \neq u_-$, we discuss two special limits of considerable interest in the literature.\\ 

\subsection{Run-and-tumble particle}

\begin{figure}[ht!]
    \centering
    \includegraphics[scale=0.75]{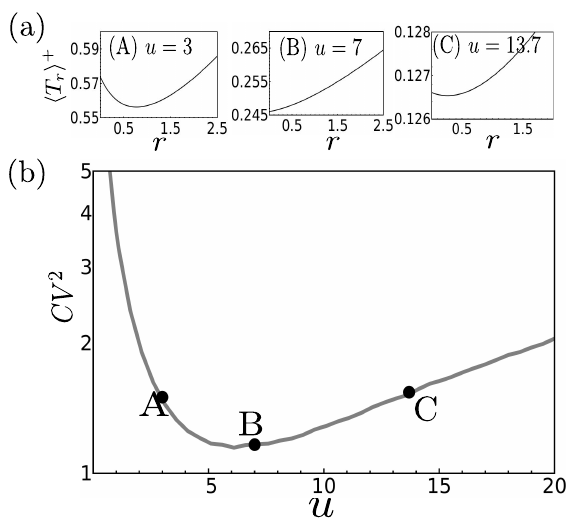}
    \caption{(a) $\langle T_r\rangle^+$ vs $r$ for three values of $u$ (low, intermediate and high) at $k_{off} = 5.5 < k_2$ (with $k_2 = 6.1)$, and $k_{on} = 20$. We see that there is a finite ORR at low and high $u$, but ORR $= 0$ at intermediate $u$.   (b) We show the $CV^2$ of first passage times for the run-and-tumble particle without resetting, at the three values of $u$ used in (a) (marked by A, B, and C). The fact that fluctuations are high  at low  and high values of $u$ (i.e. A and C), and low at intermediate value (B), justifies why resetting looses importance and again re-emerges to be advantageous as $u$ is increased.  
    }
    \label{f3}
\end{figure}

The first case is of a run-and-tumble particle for which $u_+=u_-=u$, in a chemical gradient such that run lengths along $+$ve and $-$ve $x$-axis are typically different ($k_{on}\neq k_{off}$). The MFPTs $\langle T_r\rangle^+$ and $\langle T_r\rangle^-$ given by 
Eqs.~\ref{exact_t+}, \ref{exact_t-}, now depend on four parameters $k_{on}, k_{off}, r$ and $u$. In Fig.~\ref{f2} we explore the efficacy of resetting by studying ORR in the $k_{off}-u$ parameter space. In Fig.~2(a), we study $r_*^+$ (corresponding to $\langle T_r\rangle^+$) for a fixed $k_{on}$ -- we see a region where $r_*^+=0$ (marked in gray) outside which $r_*^+\neq0$. This region has a unimodal boundary (see Fig.~\ref{f2}(b) for a zoomed view) with a maximum at $k_2$. For a fixed $k_{off}<k_2$ upon varying $u$, $r_*^+$ first vanishes, then stays zero over a range, and again becomes non-zero (see Inset I) -- thus there are two ORR vanishing transitions. The $\langle T_r\rangle^+$ versus $r$ at low $k_{off}$ for different values of $u$ are shown 
in Fig.~\ref{f3}(a). The first transition (blue line in  Fig.~\ref{f2}a) is expected due to both strong bias (large $u$) and high persistence towards the target than away from it (low $k_{off}$) \cite{Saeed_2019}. However the second transition (green line in  Fig.~\ref{f2}a) reviving the advantage of resetting at even larger $u$ is non-trivial and occurs due to the re-emergence of large FPT fluctuations (namely $CV^2 = \sigma^2/\langle t \rangle^2$) across trajectories 
(see Fig.~\ref{f3}(b)). At high $k_{off}>k_{1}$, the run lengths away from the target are longer and resetting increasingly helps -- $r_*^+$ monotonically increases with $u$ (see Inset I). This behavior changes below $k_1$. For $k_{off}\ \in (k_{1},k_2)$, $r_*^+$ has a minimum $r_{*_{\rm min}}$ (red line in Fig. 2a) with varying $u$ (see Inset I). Thus interestingly, for this intermediate range of $k_{off}$, there is an optimal speed $u$ of a run-and-tumble particle for which minimal resetting is needed for optimal target search. In inset II, we show that with decreasing $k_{on}$ (and hence decreasing residence in the attractive state) the region where $r_*^+=0$ shrinks, indicating the overall rise in benefit of resetting strategy as expected.
Finally we test the general condition of transition (Eq. (\ref{trn_condition})) by calculating the moments  $\langle T\rangle^{\sigma_f\sigma_i}$ and $\langle T^2\rangle^{\sigma_f\sigma_i}$ 
(appendix~\ref{A3}) and use those to obtain the ORR vanishing transition points (shown by filled circles in Fig.~\ref{f2}(b), which is a magnification of the $r_*^+=0$ region in Fig.~\ref{f2}(a)) -- they fall exactly on the phase boundary (made of the blue and green lines) obtained from exact calculations above. We further verify the phase boundary by obtaining the transition points numerically (shown by empty circles in Fig.~\ref{f2}(b)) using Kinetic Monte Carlo (KMC) simulations \cite{Gillespie_1977}.

The re-entrance in the advantage of resetting as function of $u$ is present for $\langle T_r\rangle^-$ as well. This is shown in the phase diagram for $r_*^-$ (corresponding to $\langle T_r\rangle^-$) in Fig.\ref{f2}(c). The filled circles correspond to the mathematical condition (in the absence of resetting) in Eq. (\ref{trn_condition}), and the blue and green lines are obtained from the exact solution of $\langle T_r\rangle^-$ (in the presence of resetting). For low values of $k_{off}$ we again see two ORR vanishing transitions as a function of $u$, indicating a re-entrance of the benefit of resetting at large $u$.\\
In summary, the origin of the phenomenon of re-entrance of advantage of resetting (Fig. \ref{f2}) is related to the $U$-shape in $CV^2$ (Fig. \ref{f3}(b)) of first passage time of target search. The latter finding is biophysically interesting -- suggesting an experimental study using bacterial mutants with different speeds $u$ to check if $CV^2$ in times of target search is optimized for a certain $u$.


\subsection{Flashing potential}

\begin{figure}[ht!]
    \centering
    \includegraphics[scale=0.25]{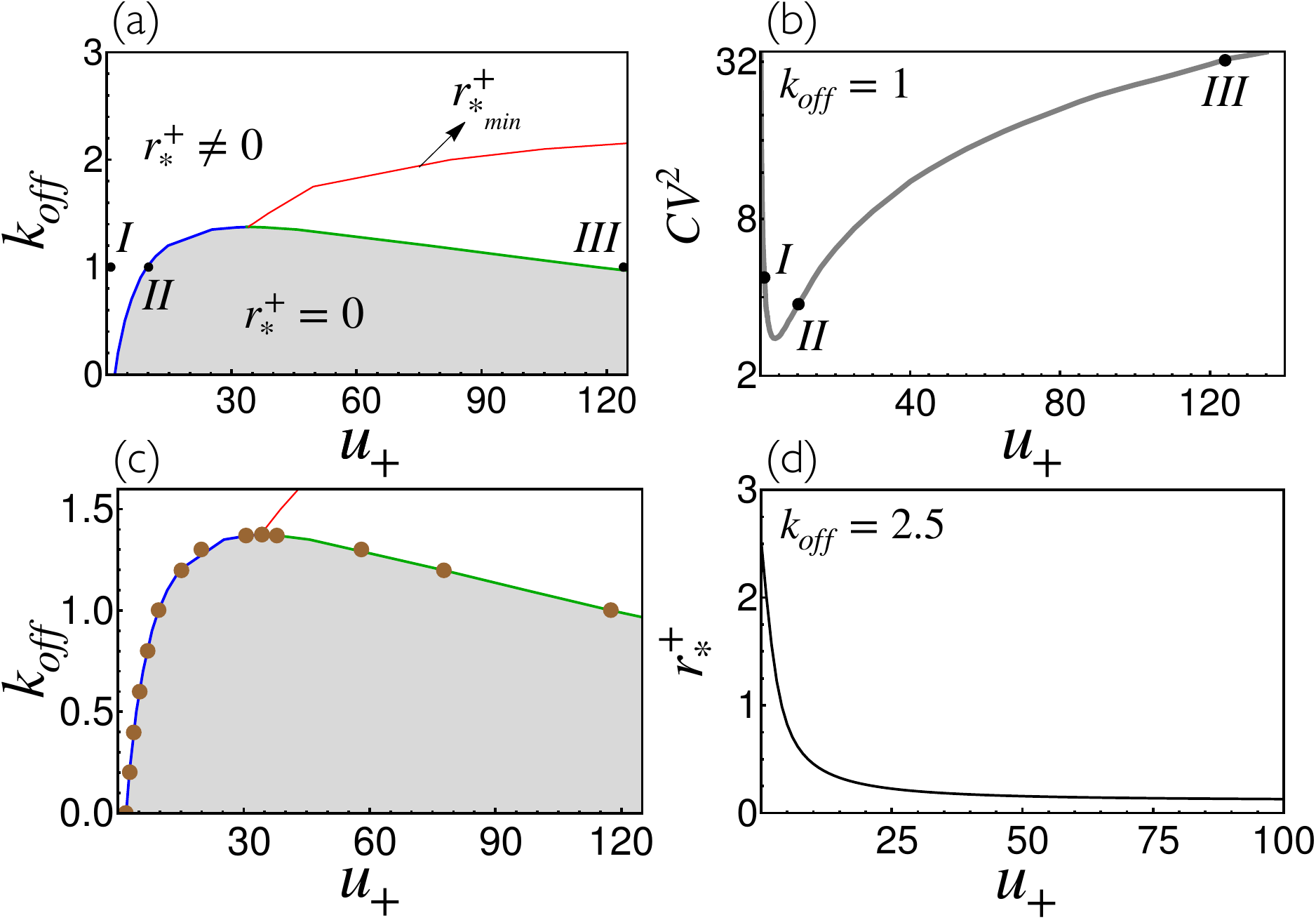}
\caption{(a) Phase diagram for $r_{*}^+$ corresponding to $\langle T_r\rangle^+$ in the $k_{off}-u$ plane for a particle in a flashing potential for $k_{on}=1$. The blue and the green lines representing the first and second transitions, determine the phase boundary between the regions $r_{*}^+=0$ (gray) and $r_{*}^{+}\neq0$ (white). The red line is the line of minimum resetting rate. All the lines are obtained using the analytic expressions for MFPT. (b) $CV^2$ as a function of $u$ with $k_{on}$ and $k_{off}$ fixed, showing the fluctuation decreases with increasing $u$ and again increases with further increase in $u$ (similar as Fig.~\ref{f3}(b)). The increasing fluctuation at higher $u$ values necessitate the re-emergence of the advantage of resetting. (c) Zoomed-in view of the phase boundary in (a). The phase boundary obtained using the mathematical condition Eq.~(\ref{Trcn}) is shown by brown filled circles. 
(d) $r_*^+$ decreases with $u_+$ for high $k_{off}$, leading to the red line of $r_{*_{\rm min}}$ in (a) moving to the right with increasing $k_{off}$, unlike the run-and-tumble case.  
}
\label{f4}
\end{figure}

Next we study a second special case of a diffusive particle subjected to an intermittently {\it flashing}  linear bias towards the target. The ORR vanishing transitions are studied by setting $u_-=0$ and  $u_+ > 0$ in the exact general results (Eqs.~\ref{exact_t+},\ref{exact_t-}). The phase diagram for $r_*^+$ (corresponding to $\langle T_r\rangle^+$) in the $k_{off}-u_+$ plane is shown in Fig.~\ref{f4}(a)
for a fixed $k_{on}$. The phase behaviour is qualitatively similar to the run-and-tumble case discussed above. The striking point is that even though the particle is not pushed away from the target in the ``$-$" state like in the run-and-tumble case, the second re-entrance 
transition occurs. The rise of FPT fluctuations across trajectories in the absence of resetting at high $u_+$ (Fig.~\ref{f4}(b)) 
ushers in again the advantage of resetting, very similarly as in run-and-tumble case. In Fig.~\ref{f4}(c)
we show the transition boundary obtained using the condition in {Eq. (\ref{trn_condition})} (filled circles), perfectly coincide with the lines obtained using the exact $\langle T_r\rangle^+$. 

This special limit (with $u_-=0$ and  $u_+ > 0$) maps to an interesting biophysical problem of   
molecular motor assisted transport inside a cell. While the idea of intermittent search in such problems has been studied earlier \cite{Benichou_2011}, a scenario with an effective resetting is discussed below.  
Denoting $\rho_b$ and $\rho_f$ as the filament-bound and freely diffusing motor protein (MP) densities, the dynamical equations for these two populations are
\beqr
\label{motor_density}
\partial_t \rho_b&=&D\partial_x^2 \rho_b+u_{+}\partial_x \rho_b+k_{on}\rho_f-k_{off}\rho_b-\gamma \rho_b+ q\delta(x-x_0), \nonumber \\ 
\partial_t \rho_f&=&D\partial_x^2 \rho_f-k_{on}\rho_f+k_{off}\rho_b-\gamma \rho_b+q\delta(x-x_0).
\eqnr
Here $u_{+}$ is the mean speed of  filament-bound MPs, $k_{on}$ and $k_{off}$ are the attachment and detachment rates to and from the filament, and $\gamma$ their degradation rate. 
The MPs are introduced at rate $q$ from a pool at a source point $x_0$ and they deliver cargo at $x=0$ (considered as a first passage).  Similar models has been used to study motor assisted cargo transport in axonal cells \cite{shee2024unc}. Typically $\gamma$ is a small quantity compared to $q$. But at low densities, if MP replenishment may balance degradation (i.e. $q = \gamma =r$), then it is {equivalent to resetting} at rate $r$ which may be tuned to minimize the time of cargo delivery. Our result in this section suggests that at low and high processive speeds $u_+$, this resetting protocol would be optimal which implies possibility of optimizing cargo delivery times by tuning replenishment and degradation rates. For intermediate $u_+$ such tuning would not help.\\

\subsection{General case with unequal biases}
\begin{figure}[ht!]
\centering
\includegraphics[scale=0.45]{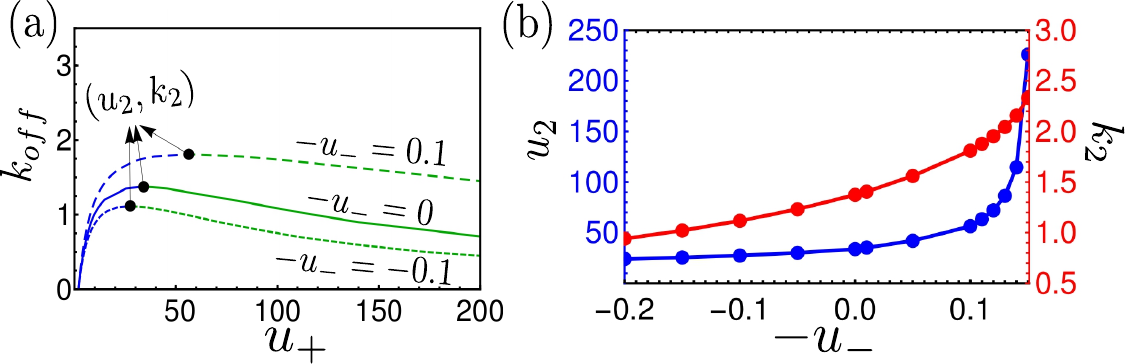}
\caption{
For arbitrary linear potentials, we show the re-entrance transition in ORR for $k_{on}=1$. (a) The two transitions lines in $k_{off}-u_+$ plane in blue and green meet at a finite point $(u_2,k_2)$ for three different (repulsive, unbiased and attractive) strengths of $-u_-$. (b) Systematic variation of $u_2$ and $k_2$ extracted from curves as shown in (a), plotted as a function of $-u_-$.  As the attractive strength increase, $u_2$ tends to diverge, implying the disappearance of re-entrance beyond a point.
}
\label{f5}
\end{figure}

Finally we return to the general case of arbitrary linear potentials. We plot in Fig.\ref{f5}(a), the ORR $r_*^+$ vanishing boundaries as a function of the strength of the attractive bias $u_+$ 
for different values of the potential strength $u_-$ (of the ``$-$" state). Quite strikingly, the re-entrance occurs even for weakly attractive states ($-u_- > 0$ but small), with the blue and green lines meeting at a finite point ($u_2,k_2$). This latter point approaches infinity ($u_2$ tends to diverge) as the strength of attraction $-u_-$ increases further (see Fig.\ref{f5}b). Thus for high enough $-u_- > 0$ there would be a single transition (only blue line), as the joint attractive biases of both ``$+$'' and ``$-$'' states is expected to  annul the advantage of resetting at once. Interestingly, linear potentials with unequal and non-zero $u_+$ and $u_-$ is reminiscent of dynamic microtubules \cite{Dogterom_1993,Aparna2017}; here  $u_+$ denotes the polymerization rate of tubulin dimers. While a possible way of resetting for $x>x_0$ could be by chopping filaments using laser ablation, however resetting lengths $x<x_0$ seems challenging experimentally. Studies in future may explore such one-sided resetting.\\

\section{Non-linear potentials}
\begin{figure}[ht!]
\centering
\hspace{-0.1cm}\includegraphics[scale=0.5]{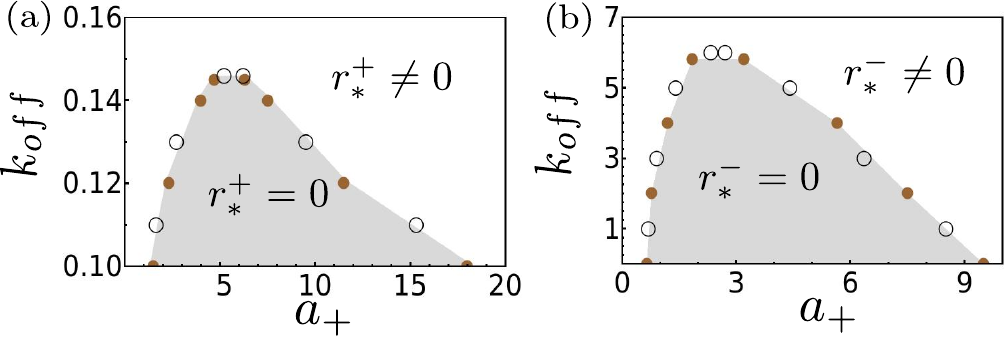}
\caption{ORR vanishing transition boundaries obtained using Eq.~\ref{trn_condition} (filled circles) and direct simulations (empty circles) for (a) quadratic flashing trap, corresponding to $\langle T_r\rangle^+$ for $k_{on}=0.3$, and for (b) quadratic trap and anti-trap, corresponding to $\langle T_r\rangle^-$ for $k_{on}=20$.
}
\label{f6}
\end{figure}

So far we have discussed toggling under linear potentials, while nonlinear potentials are common and realizable in experiments, for example quadratic potentials \cite{Friedman_2020,Besga_2020} and ratchets \cite{ethier_2018}. With non-linear potentials $V_+$ and $V_-$ the exact solution of $\tilde{Q}_r^{\sigma_f \sigma_i}(s)$ or MFPTs with state toggling are challenging. Yet the general condition (Eq. (\ref{trn_condition})) should be applicable in these cases as well. To test this, we study two cases: (a) a flashing quadratic trap with $V_{+}=a_{+}x^2$, $V_{-}=0$, and (b) with $V_{+}=a_{+}x^2$ (trap), $V_{-}=-V_{+}$ (anti-trap). We perform simulations to obtain the conditional probabilities ${Q}_0^{\sigma_f \sigma_i}(t)$ (see details of the procedure in appendix~\ref{A4})
and numerically integrate those to obtain the required moments ${\langle T^k \rangle}^{\sigma_f \sigma_i}$ -- these moments are then used in the condition Eq. (\ref{trn_condition}) to locate the ORR transitions (marked by filled circles in Fig.~\ref{f6}). We also directly obtained the ORR vanishing from simulations (appendix~\ref{A4})
of the corresponding problems with resetting which verify the boundaries (marked by open circles in Fig.~\ref{f6}). In both the cases (Fig.~\ref{f6}(a) and (b)) we see a re-entrant transition as a function of the trap stiffness $a_+$.


\section{Conclusion}

Stochastic processes with state toggling are quite common -- some examples we discussed are run-and-tumble particle, molecular motors switching between a directed walk on a filament to diffusive transport in cytoplasm, dynamic instability of microtubules, and flashing optical traps. If such problems involve target search, a natural question arises whether stochastic resetting would enhance the efficacy of the search process. Here we derived a general mathematical condition which determines the boundary in the parameter space indicating a transition between \emph{beneficial} and \emph{harmful} resetting strategy. 

For problems with state toggling under arbitrary linear bias, we have exactly solved both the MFPT and the general mathematical condition for ORR transition. These general solutions for asymmetric toggling rates, shed new light on the problem of diffusive run-and-tumble particle in a chemical gradient. We find that there is a re-entrance of resetting advantage with increasing speed of the particle. Moreover, we demonstrated that the phenomenon is very generic -- it is true when an attractive state switches to a repulsive, neutral, or even another weakly attractive state. We showed that it persists for non-linear potentials like quadratic flashing traps. While the loss of advantage of resetting under bias has been extensively studied in the past, the re-emergence of its importance at even higher bias is a bit unexpected. This generic feature arises as first passage time-scale fluctuations become high under state toggling and strong bias, and the search time is optimized by resetting. 

The theoretical problem studied is relevant to a broad range of experimental systems; we hope our predictions will be tested in future experiments under appropriate conditions. Although we have restricted our examples to diffusive transport problems, the general mathematical condition derived here may be used to study biochemical reactions too which involve state toggling and target search -- a potential theoretical direction which may be pursued in future.\\

\noindent
\section{Acknowledgement}
 AN acknowledges Science and Engineering Research Board (SERB), India (Project No. MTR/2023/000507) for financial support. DD and AN thanks the Max-Planck Institute for the Physics of Complex Systems (MPIPKS), Dresden, for hospitality and support during summer visit in 2024. HKB thanks UGC, India for the Junior Research Fellowship. \\

\appendix\label{A}

\section{Derivation of the General condition of ORR vanishing transition}\label{A1}


From Eq.~(\ref{Lap_renewal}) we get the following four explicit equations: 
\begin{eqnarray}
    \begin{bmatrix}
    \Tilde{Q}_r^{++}(s)\\\Tilde{Q}_r^{-+}(s)
    \end{bmatrix} =\begin{bmatrix}
    \Tilde{Q}_0^{++}(r+s)\\\Tilde{Q}_0^{-+}(r+s)
    \end{bmatrix}+A
    \begin{bmatrix}
        \Tilde{Q}_r^{++}(s)\\ \Tilde{Q}_r^{-+}(s)
    \end{bmatrix} ,\nonumber\\ \nonumber\\
    \begin{bmatrix}
    \Tilde{Q}_r^{+-}(s)\\\Tilde{Q}_r^{--}(s)
    \end{bmatrix} =
    \begin{bmatrix}
    \Tilde{Q}_0^{+-}(r+s)\\\Tilde{Q}_0^{--}(r+s)
    \end{bmatrix}+A
    \begin{bmatrix}
        \Tilde{Q}_r^{+-}(s)\\ \Tilde{Q}_r^{--}(s)
    \end{bmatrix},\label{explicit_eqs}
\end{eqnarray}
where
\begin{eqnarray}
    A=
    \begin{bmatrix}
    r\Tilde{Q}_0^{++}(r+s)& r\Tilde{Q}_0^{+-}(r+s)\\
    r\Tilde{Q}_0^{-+}(r+s) & r\Tilde{Q}_0^{--}(r+s)
    \end{bmatrix}.\nonumber
\end{eqnarray}
Solving the algebraic equations for the quantities  $\Tilde{Q}_r^{\sigma_f\sigma_i}(s)$, we get
\begin{widetext}
\begin{eqnarray}
   \begin{bmatrix}
    \Tilde{Q}_r^{++}(s)\\\\\Tilde{Q}_r^{-+}(s)\\\\\Tilde{Q}_r^{+-}(s)\\\\\Tilde{Q}_r^{--}(s)
    \end{bmatrix} =\frac{1}{det(I-A)}
    \begin{bmatrix}
        \Tilde{Q}_0^{++}(r+s)-r\Tilde{Q}_0^{--}(r+s)\Tilde{Q}_0^{++}(r+s) +r\Tilde{Q}_0^{+-}(r+s)\Tilde{Q}_0^{-+}(r+s)\\\\
        \Tilde{Q}_0^{-+}(r+s)\\\\
        \Tilde{Q}_0^{+-}(r+s)\\\\
        \Tilde{Q}_0^{--}(r+s)-r\Tilde{Q}_0^{++}(r+s)\Tilde{Q}_0^{--}(r+s)+r\Tilde{Q}_0^{-+}(r+s)\Tilde{Q}_0^{+-}(r+s)
    \end{bmatrix}, \label{Qr(s)}
\end{eqnarray}
\end{widetext}
where $I$ denotes the $2\times 2$ identity matrix.\\

From the above equation we can get the survival probabilities in Laplace space in the presence of resetting $\Tilde{Q}_r^{+}(s)=\Tilde{Q}_r^{++}(s)+\Tilde{Q}_r^{-+}(s)$ and $\Tilde{Q}_r^{-}(s)=\Tilde{Q}_r^{+-}(s)+\Tilde{Q}_r^{--}(s)$
in terms of quantities in the absence of resetting. Then taking the limit $s\rightarrow 0$ we obtain the two MFPTs $\langle T_r \rangle^+=\Tilde{Q}_r^+(s\rightarrow0)$ and $\langle T_r \rangle^-=\Tilde{Q}_r^-(s\rightarrow0)$ starting from the state ($+,x_0$) and ($-,x_0$) respectively (see Eq.~(\ref{Tasr}) in the main text).\\

We define the two ORRs $r_*^+$ and $r_*^-$ using  $\frac{d \langle T_r \rangle^+ }{dr}\big |_{r_*^+}=0$ and $\frac{d \langle T_r \rangle^- }{dr}\big |_{r_*^-}=0$. As we are interested in the ORR vanishing transition, we are interested in the limit $r_*^\pm \rightarrow 0$. Hence, Taylor expanding $\langle T_r \rangle^+=a_0^++ra_1^++O(r^2)$  and $\langle T_r \rangle^-=a_0^-+ra_1^-+O(r^2)$, the conditions for ORR vanishing transitions are as follows:
\begin{eqnarray}
    a_1^+=\big( \langle T \rangle ^{++}\big)^2&&+ \langle T \rangle ^{-+}(\langle T \rangle ^{+-} + \langle T \rangle ^{++}+ \langle T \rangle ^{--})\nonumber\\
    &&-\frac{1}{2}( \langle T^2 \rangle ^{++}+ \langle T^2 \rangle ^{-+})=0 ,\nonumber\\\nonumber\\\nonumber
    a_1^-=\big( \langle T \rangle ^{--}\big)^2&&+ \langle T \rangle ^{+-}(\langle T \rangle ^{-+}+ \langle T \rangle ^{++} + \langle T \rangle ^{--})\nonumber\\
    &&-\frac{1}{2}( \langle T^2 \rangle ^{+-}+ \langle T^2 \rangle ^{--})=0.\nonumber\\\label{Trcn}
\end{eqnarray}
In the above the symbols $\langle T\rangle^{\sigma_f\sigma_i}=\Tilde{Q}_0^{\sigma_f\sigma_i}(r\rightarrow 0)$ and $\langle T^2\rangle^{\sigma_f\sigma_i}=-2\frac{\partial \Tilde{Q}_0^{\sigma_f\sigma_i}(r)}{\partial r}\Big|_{r\rightarrow 0}$. These two conditions can be combined to write a formal single condition 
(see Eq.~(\ref{trn_condition}) in the main text)
\begin{eqnarray}
\label{ORR_cond_A}
    \big(\langle T \rangle ^{\sigma_i\sigma_i}\big)^2&&+ \langle T \rangle ^{-\sigma_i\sigma_i}\Big( \langle T \rangle ^{\sigma_i-\sigma_i}+ \langle T \rangle ^{-\sigma_i-\sigma_i}+ \langle T \rangle ^{\sigma_i\sigma_i}\Big)\nonumber\\
    &&-\frac{1}{2}\Big( \langle T^2 \rangle ^{\sigma_i\sigma_i}+ \langle T^2 \rangle ^{-\sigma_i\sigma_i}\Big)=0.\label{trn_condition_A}
\end{eqnarray}
\section{Toggling between two states with linear bias: Solution of Survival probabilities in Laplace space, and MFPTs, in the presence of resetting}
\label{A2}
In terms of non-dimensionalized variables and parameters, Eq.~(\ref{survival_general_0}) can be written as
  \begin{eqnarray}
  \frac{\partial Q_r^+(x,t)}{\partial t}&=&\frac{\partial^2 Q_r^+(x,t)}{\partial x^2}-u_+\frac{\partial Q_r^+(x,t)}{\partial x}+rQ_r^+(1,t)\nonumber\\
  &&-rQ_r^+(x,t)+k_{off}\left[Q_r^-(x,t)-Q_r^+(x,t)\right],\nonumber\\\nonumber\\
  \frac{\partial Q_r^-(x,t)}{\partial t}&=&\frac{\partial^2 Q_r^-(x,t)}{\partial x^2}+u_-\frac{\partial Q_r^-(x,t)}{\partial x}+rQ_r^-(1,t)\nonumber\\
  &&-rQ_r^-(x,t)+k_{on}\left[Q_r^+(x,t)-Q_r^-(x,t)\right],\nonumber\\
   \label{survival_general_1}
\end{eqnarray}
where the rescaling used as: $t D/x_0^2\rightarrow t$, ~ $x/x_0\rightarrow x$, ~ $x_0v_{\pm}/D\rightarrow u_\pm$, ~
$k_{off}x_0^2/D\rightarrow k_{off}$, ~ $k_{on}x_0^2/D\rightarrow k_{on}$, ~
$rx_0^2/D\rightarrow r$.
Note that due to rescaling of space, $x=x_0$ has become $x=1$.

Taking Laplace transformation of the Eq. (\ref{survival_general_1}) corresponding to the time domain, and using the  initial conditions, we have two linearly coupled inhomogeneous ODEs:
\begin{eqnarray}
s\tilde{Q}_r^+(x,s)-1&=&\frac{\partial^2 \tilde{Q}_r^+(x,s)}{\partial x^2}-u_+\frac{\partial \tilde{Q}_r^+(x,s)}{\partial x}+r\tilde{Q}_r^+(1,s)\nonumber\\
&&-r\tilde{Q}_r^+(x,s)+k_{off}\left[\tilde{Q}_r^-(x,s)-\tilde{Q}_r^+(x,s)\right], \ \ \nonumber\\\nonumber\\
s\tilde{Q}_r^-(x,s)-1&=&\frac{\partial^2 \tilde{Q}_r^-(x,s)}{\partial x^2}+u_-\frac{\partial \tilde{Q}_r^-(x,s)}{\partial x}+r\tilde{Q}_r^-(1,s)\nonumber\\
&&-r\tilde{Q}_r^-(x,s)+k_{on}\left[\tilde{Q}_r^+(x,s)-\tilde{Q}_r^-(x,s)\right]. \nonumber\\ \label{Laplace_general}
\end{eqnarray}
By shifting the above functions as $\Tilde{Q}_r^+(x,s)=y^+(x,s)-a$ and $\Tilde{Q}_r^-(x,s)=y^-(x,s)-b$ and choosing $a$ and $b$ to be 
\begin{eqnarray}
    a&=&-\frac{1}{r+s}-\frac{r(r+s+k_{on})\Tilde{Q}_r^+(1,s)+rk_{off} \Tilde{Q}_r^-(1,s)}{(r+s) (r+s+k_{on}+k_{off})},\nonumber\\
    b&=&-\frac{1}{r+s}-\frac{  rk_{on} \Tilde{Q}_r^+(1,s)  +(r + s+k_{off}) \Tilde{Q}_r^-r(1,s)} {(r + s) (k_{off} + k_{on} + r + s)},\nonumber\\ \label{general_constants}
\end{eqnarray}
we obtain two homogeneous ODEs for $y^\pm(x,s)$ as follows:
\begin{align}
&\frac{\partial^2 y^+}{\partial x^2}-u_+\frac{\partial y^+}{\partial{x}}-\left(r+s+k_{off}\right)y^+ + k_{off}y^-=0,\notag\\
&\frac{\partial^2 y^-}{\partial x^2}+u_-\frac{\partial y^-}{\partial{x}}-\left(r+s+k_{on}\right)y^- + k_{on} y^+=0.\label{general_inhomo}
\end{align}
To solve these equations, we substitute an ansatz $y^{\pm}(x,t)\sim A_\pm e^{-\lambda x}$ in Eq.(\ref{general_inhomo}). For a non trivial $A_\pm$, the determinant of the matrix multiplying the vector $\begin{bmatrix}
    A_+\\A_-
\end{bmatrix}$ should be zero, which leads to the following quartic equation for $\lambda:$
\begin{eqnarray}
   \lambda^4+(u_+&-&u_-)\lambda^3-\left[2(r+s)+k_{off}+k_{on}+u_+u_-\right]\lambda^2\nonumber\\
   &+&\left[(r+s+k_{off})u_--(r+s+k_{on})u_+\right]\lambda\nonumber\\ & & +(r+s)(r+s+k_{on}+k_{off})=0.\label{quartic}
\end{eqnarray}
To ensure that $y^\pm(x,s)$ are finite as $x\rightarrow\infty$, only positive roots of the above equation are physically allowed. There are two such positive roots which we denote by $\lambda_1$ and $\lambda_2$ and the general solutions of $y^\pm$ are:
\begin{eqnarray}
	y^+(x,s)&=&A_1e^{-\lambda_1 x}+A_2e^{-\lambda_2 x},\notag\\
	y^-(x,s)&=&B_1e^{-\lambda_1 x}+B_2e^{-\lambda_2 x},\label{ypm}
\end{eqnarray}
where $A_1,A_2,B_1$ and $B_2$ are constants. Only two of them are independent as we can see by substituting solutions (\ref{ypm}) in one of the Eq. (\ref{general_inhomo}), and setting the coefficients of $x$-dependent functions to zero.  We get
\begin{eqnarray}
    B_1&=&-\frac{1}{k_{off}}\left[ \lambda_1^2+\lambda_1u_+-\left(r+s+k_{off} \right) \right]A_1,\notag\\
    B_2&=&-\frac{1}{k_{off}}\left[ \lambda_2^2+\lambda_2u_+-\left( r+s+k_{off} \right) \right]A_2.
\end{eqnarray}
Substituting these relations in Eq. (\ref{ypm}), we may obtain $y^\pm(x,s)$ and hence $\Tilde{Q}_r^\pm(x,s)$ (see Eq. \ref{general_constants}): 
\begin{eqnarray}
    \Tilde{Q}_r^+(x,s)&=&A_1e^{-\lambda_1 x}+A_2e^{-\lambda_2 x}-a,\nonumber\\
    \Tilde{Q}_r^-(x,s)&=&-\frac{1}{k_{on}}\Big\{\left[ \lambda_1^2-\lambda_1u_--\left( r+s+k_{on}\right) \right]A_1e^{-\lambda_1 x}\nonumber\\
    &&+\left[ \lambda_2^2-\lambda_2u_--\left( r+s+k_{on} \right) \right]A_2e^{-\lambda_2 x}\Big\} -b.\nonumber\\ \label{General_sol2}
\end{eqnarray}

The two unknown constants $A_1$ and $A_2$ in Eq. (\ref{General_sol2}) can be solved by using the boundary conditions at $x=0$ ($Q_r^\pm (0,t)=0$). This leads to 
\begin{eqnarray}
    A_1&=&\frac{
    \begin{array}{l}
    -(r + s)(r+s+k_{on}+k_{off})\big(1+r\Tilde{Q}_r^+(1,s)\big)+ \lambda_2(u_+\\
    +\lambda_2)\big[k_{off}\big(1+ r\Tilde{Q}_r^-(1,s)\big) + (k_{on} + r + s)\big(1 + r\Tilde{Q}_r^+(1,s)\big)\big]
    \end{array}
    }{
    (r+s)(r+s+k_{on}+k_{off})(\lambda_1-\lambda_2)(u_++\lambda_1+\lambda_2)
    },\nonumber\\
    A_2 &=& \frac{
\begin{array}{l}
(r + s)(r + s + k_{on} + k_{off})\big(1 + r \Tilde{Q}_r^+(1, s)\big)- \lambda_1 (u_+ \\
+ \lambda_1) \big[ k_{off}\big(1 + r \Tilde{Q}_r^-(1, s)\big)
+ (k_{on} + r + s)\big(1 + r \Tilde{Q}_r^+(1, s)\big) \big]
\end{array}
}{
(r + s)(r + s + k_{on} + k_{off})(\lambda_1 - \lambda_2)
(u_+ + \lambda_1 + \lambda_2)
}.\nonumber\\
\end{eqnarray}

 Note that the constants $A_1,A_2$ linearly depend on $\Tilde{Q}_r^\pm(1,s)$. Hence putting these equations back in Eq. (\ref{General_sol2}) and setting $x=1$ on the left hand side, one may linearly solve for $\Tilde{Q}_r^\pm(1,s)$ from Eq. (\ref{General_sol2}). The expressions of $\Tilde{Q}_r^\pm(1,s)$ and $\lambda_1,\lambda_2$ are very lengthy to be written here --further analysis of those expressions for the main manuscript have been done using \textbf{Mathematica}. The expressions of the mean first passage times (MFPTs) $\langle T_r\rangle^\pm$, are obtained using $\Tilde{Q}_r^\pm(1,s\rightarrow 0)$ as

\begin{widetext}
\begin{eqnarray}
\label{exact_t+}
\langle T_r\rangle^+ &=&\Bigg [ \mu _ 1^2\Big(e^{\mu _ 2} - 1  \Big) \Big(e^{\mu _ 1}(k_ {off} + k_ {on} ) + r \Big) + r\Big(e^{\mu _ 1} -   e^{\mu _ 2}  \Big) \Big(k_ {off}+ k_ {on} +  r \Big)  +  \mu _ 1  u_+\Big(e^{\mu _ 2} - 1  \Big)  \Big(e^{\mu _ 1} (k_ {off} +  k_ {on}  )\nonumber\\
& +& r  \Big) -  \mu _ 2\Big(e^{\mu _ 1} -  1  \Big) \Big(\mu _ 2 + u_+  \Big) \Big(e^{\mu _ 2} (k_ {off} +  k_ {on}  ) + r  \Big) \Bigg] \Bigg{/} \Bigg [r \bigg(-e^{\mu _ 2} \Big(\mu _ 2 (\mu _ 2 + \ u_+  ) - r  \Big) \Big(k_ {off} + k_ {on} +   r - \mu _ 1\nonumber\\
& \times&(\mu _ 1+ u_+  )  \Big) + e^{\mu _ 1} \Big(\mu _ 1 (\mu _ 1 + u_+  ) -   r  \Big) \Big(k_ {off} + k_ {on}   + r-\mu _ 2 (\mu _ 2 + u_+  )  \Big)   +  r\Big(\mu _ 1 - \mu _ 2  \Big)  \Big(\mu _ 1 + \mu _ 2 + u_+  \Big)  \bigg)\ \Bigg ].\nonumber\\ \\
\langle T_r\rangle^-&=& \Bigg [ e^{\mu _2} \bigg( k_{off} ^2 \Big(\mu _2 (\mu _2+ u_+ )-r \Big)+ k_{off} \Big( k_{on} \big(\mu _2 (\mu _2 +u_+ ) -r \big)+ \big(\mu _1 (\mu _1+ u_+ )-r \big) \big(2 r-\mu _2 (\mu _2+   u_+ ) \big) \Big)\nonumber\\ 
&+& \Big( k_{on} +r\Big) \Big(r-\mu _1 (\mu _1+ u_+ ) \Big) \Big(\mu _2 (\mu _2+ u_+ )-r \Big) \bigg)+ e^{\mu _1} \bigg( k_{off} ^2 \Big(r-\mu _1 (\mu _1+ u_+ ) \Big)+ k_{off} \Big( k_{on} \big(r-\mu _1 (\mu _1\nonumber\\
&+& u_+ ) \big) + \big(2 r-\mu _1 (\mu _1 + u_+ ) \big) \big(r-\mu _2 (\mu _2+  u_+ ) \big) \Big) +\Big( k_{on} +r\Big) \Big(r-\mu _1 (\mu _1+ u_+ ) \Big) \Big(r-\mu _2 (\mu _2+ u_+  ) \Big) \bigg)\nonumber\\
&+& k_{off} e^{\mu _1+\mu _2} (\mu _1-\mu _2 ) ( k_{off} + k_{on} ) (\mu _1+\mu _2+ u_+ )- r k_{off} (\mu _1-\mu _2 )(\mu _1+\mu _2+ u_+ ) \Bigg ] \bigg /\nonumber\\ 
&\Bigg [r& k_{off}   \bigg(-e^{\mu _2} \Big(\mu _2 (\mu _2 + u_+ ) -r \Big) \Big( k_{off} + k_{on} +r-\mu _1    (\mu _1+   u_+  )  \Big)+e^{\mu _1}    \Big(\mu _1    (\mu _1 +   u_+  )- r  \Big) \Big( k_{off} + k_{on}\nonumber\\
&+&r-\mu _2    (\mu _2+   u_+  )  \Big)+   r(\mu _1
- \mu _2  )  (\mu _1+ \mu _2 +   u_+  )  \bigg)\Bigg ].
\label{exact_t-}
\end{eqnarray}
\end{widetext}

{Here $\mu_{1,2}=\lambda_{1,2}(s\rightarrow 0)$. Eqs.(\ref{exact_t+}-\ref{exact_t-}) give the exact MFPTs which may be used to study ORR for any values of the linear potentials $V_{\pm}(x)=\pm u_{\pm} x$ as done in the main text.}


\section{Toggling between two states with linear bias: Solution of the joint survival probabilities in Laplace space, in the absence of resetting}\label{A3}
Taking Laplace transform of Eq.~(\ref{gen_wr}), and using the initial conditions,  we have four linearly coupled inhomogeneous ODEs:

\begin{eqnarray}
    s\Tilde{Q}_0^{++}(x,s)-1= \frac{d^2 \Tilde{Q}_0^{++}(x,s)}{d x^2}&-&u_+\frac{d\Tilde{Q}_0^{++}(x,s)}{dx}\nonumber\\
    +\ &k_{off}&\big[\Tilde{Q}_0^{+-}(x,s)-\Tilde{Q}_0^{++}(x,s)\big],\nonumber\\
    s\Tilde{Q}_0^{-+}(x,s)= \frac{d^2 \Tilde{Q}_0^{-+}(x,s)}{d x^2}&-&u_+\frac{d\Tilde{Q}_0^{-+}(x,s)}{dx}\nonumber\\
    +\ &k_{off}&\big[\Tilde{Q}_0^{--}(x,s)-\Tilde{Q}_0^{-+}(x,s)\big],\nonumber\\
    s\Tilde{Q}_0^{+-}(x,s)= \frac{d^2 \Tilde{Q}_0^{+-}(x,s)}{d x^2}&+&u_-\frac{d\Tilde{Q}_0^{+-}(x,s)}{dx}\nonumber\\
    +\ &k_{on}&\big[\Tilde{Q}_0^{++}(x,s)-\Tilde{Q}_0^{+-}(x,s)\big],\nonumber\\
    s\Tilde{Q}_0^{--}(x,s)-1= \frac{d^2 \Tilde{Q}_0^{--}(x,s)}{d x^2}&+&u_-\frac{d\Tilde{Q}_0^{--}(x,s)}{dx}\nonumber\\
    +\ &k_{on}&\big[\Tilde{Q}_0^{-+}(x,s)-\Tilde{Q}_0^{--}(x,s)\big].\nonumber\\ \label{laplace_gen4}
\end{eqnarray}
By shifting the joint probabilities as 
$\Tilde{Q}_0^{++}(x,s)=y^{++}(x,s)-a$,
$\Tilde{Q}_0^{-+}(x,s)=y^{-+}(x,s)-b$,
$\Tilde{Q}_0^{+-}(x,s)=y^{+-}(x,s)-c$ and $\Tilde{Q}_0^{--}(x,s)=y^{--}(x,s)-d$ we obtain four ODEs for the new functions. Since, these equations are coupled in pairs we write them as follows:
\begin{eqnarray}
    &&\frac{d^2y^{++}}{dx^2}-u_+\frac{d y^{++}}{dx}-(s+k_{off})y^{++}+k_{off}y^{+-}=0,\nonumber\\  
    &&\frac{d^2y^{+-}}{dx^2}+u_-\frac{d y^{+-}}{dx}-(s+k_{on})y^{+-}+k_{on}y^{++}=0,\nonumber\\
   \label{y1}
\end{eqnarray}
\begin{eqnarray}
    &&\frac{d^2y^{-+}}{dx^2}-u_+\frac{d y^{-+}}{dx}-(s+k_{off})y^{-+}+k_{off}y^{--}=0,\nonumber\\
     &&\frac{d^2y^{--}}{dx^2}+u_-\frac{d y^{--}}{dx}-(s+k_{on})y^{--}+k_{on}y^{-+}=0,\nonumber\\ \label{y2}
\end{eqnarray}
where $a$, $b$, $c$ and $d$ are chosen as
\begin{eqnarray}
    a&=&-\frac{k_{on}+s}{s (k_{on}+k_{off}+s)},\nonumber\\
    b&=&-\frac{k_{off}}{s (k_{on}+k_{off}+s)},\nonumber\\
    c&=&-\frac{k_{on}}{s (k_{on}+k_{off}+s)},\nonumber\\
    d&=&-\frac{k_{off}+s}{s (k_{on}+k_{off}+s)}.
\end{eqnarray} 
In order to solve the equations, we first substitute an ansatz $y^{+\pm}(x,t)\sim A_{+\pm} e^{-\lambda x}$ in Eq.(\ref{y1}).
For non trivial $A_{+\pm}$, the determinant of the matrix multiplying the vector $\begin{bmatrix}
    A_{++}\\A_{+-}
\end{bmatrix}$ should be zero, which leads to the following quartic equation for $\lambda:$
\begin{eqnarray}
 \lambda^4&+&(u_+-u_-)\lambda^3-\big[u_+u_-+(2s+k_{on}+k_{off})\big]\lambda^2\nonumber\\
 &-&\big[(s+k_{on})u_+-(s+k_{off})u_-\big]\lambda+s(s+k_{on}+k_{off})=0.\nonumber\\\label{8eq}
\end{eqnarray}
Substituting a similar ansatz $y^{-\pm}(x,t)\sim A_{-\pm} e^{-\lambda x}$ in Eq. (\ref{y2}), and following similar procedure as above, we obtain the same equation (Eq. (\ref{8eq})) for $\lambda$.

For finite $y^{\pm\pm}$ at $x\rightarrow \infty$, only positive roots are allowed for physically relevant solution. Eq. (\ref{8eq}) has two distinct positive real roots which we denote by $\lambda_1$ and $\lambda_2$. Thus the general solutions of $y^{\pm\pm}$ are:
\begin{eqnarray}
    y^{++}(x,s)&=&A_1e^{-\lambda_1 x}+A_2e^{-\lambda_2 x},\nonumber\\
    y^{+-}(x,s)&=&C_1e^{-\lambda_1 x}+C_2e^{-\lambda_2 x},\nonumber\\
    y^{-+}(x,s)&=&B_1e^{-\lambda_1 x}+B_2e^{-\lambda_2 x},\nonumber\\
    y^{--}(x,s)&=&D_1e^{-\lambda_1 x}+D_2e^{-\lambda_2 x}.\label{gensol}
\end{eqnarray}
We have eight unknown constants $A_{1,2}$, $B_{1,2}$, $C_{1,2}$ and $D_{1,2}$. Substituting Eq. (\ref{gensol}) in Eq. (\ref{y1}) and Eq. (\ref{y2}), and comparing the coefficients of the exponential functions, we have only four independent unknown coefficients, while others are dependent as follows:
\begin{eqnarray}
    C_1&=-\frac{1}{k_{off}}\left[ \lambda_1^2+\lambda_1u_+-\left(s+k_{off} \right) \right]A_1,\notag\\
    C_2&=-\frac{1}{k_{off}}\left[ \lambda_2^2+\lambda_2u_+-\left( s+k_{off} \right) \right]A_2,\notag\\
    D_1&=-\frac{1}{k_{off}}\left[ \lambda_1^2+\lambda_1u_+-\left(s+k_{off} \right) \right]B_1,\notag\\
    D_2&=-\frac{1}{k_{off}}\left[ \lambda_2^2+\lambda_2u_+-\left( s+k_{off} \right) \right]B_2.\label{relations}
\end{eqnarray}
Using Eq. (\ref{relations}) we obtain $y^{\pm\pm}$ and hence $\Tilde{Q}_0^{\pm\pm}(x,s)$, still involving four unknown constants $A_{1,2}$, $B_{1,2}$:
\begin{eqnarray}
    \Tilde{Q}_0^{++}(x,s)&=&A_1e^{-\lambda_1 x}+A_2e^{-\lambda_2 x}-a,\nonumber\\
    \Tilde{Q}_0^{-+}(x,s)&=&B_1e^{-\lambda_1 x}+B_2e^{-\lambda_2 x}-b,\nonumber\\
    \Tilde{Q}_0^{+-}(x,s)&=&-\frac{1}{k_{off}}\Big\{[ \lambda_1^2+\lambda_1u_+-(s+k_{off} )]A_1e^{-\lambda_1 x}\nonumber\\
    &+&[ \lambda_2^2+\lambda_2u_+-( s+k_{off} )A_2e^{-\lambda_2 x}\Big\}-c,\nonumber\\
    \Tilde{Q}_0^{--}(x,s)&=&-\frac{1}{k_{off}}\Big\{[ \lambda_1^2+\lambda_1u_+-(s+k_{off} ) ]B_1e^{-\lambda_1 x}\nonumber\\
    &+&[ \lambda_2^2+\lambda_2u_+-( s+k_{off} ) ]B_2e^{-\lambda_2 x}\Big\}-d.\nonumber\\ \label{QS}
\end{eqnarray}
Finally, the unknown constants are calculated by using  the four boundary conditions ($Q_0^{\pm\pm}(0,t)=0$), which imply $\Tilde{Q}_0^{\pm\pm}(0,s)=0$. Substituting in Eq. (\ref{QS}) we get:
\begin{eqnarray}
    A_1&=&-\frac{k_{off} s+(k_{on}+s) (s-\lambda_2 (\lambda_2+u_+))}{s (\lambda_1-\lambda_2) (k_{off}+k_{on}+s) (\lambda_1+\lambda_2+u_+)},\nonumber\\
    A_2&=&\frac{k_{off} s+(k_{on}+s) (s-\lambda_1 (\lambda_1+u_+))}{s (\lambda_1-\lambda_2) (k_{off}+k_{on}+s) (\lambda_1+\lambda_2+u_+)},\nonumber\\
    B_1&=&\frac{k_{off} \lambda_2 (u_+ + \lambda_2)}{s (k_{off} + k_{on} + 
     s) (\lambda_1 - \lambda_2) (u_+ + \lambda_1 + \lambda_2)},\nonumber\\
     B_2&=&-\frac{\lambda_1 k_{off} (\lambda_1+u_+)}{s (\lambda_1-\lambda_2) (k_{off}+k_{on}+s) (\lambda_1+\lambda_2+u_+)}.\nonumber\\
\end{eqnarray}
Taking the limit $s\rightarrow 0$, we obtain the relevant moments $\langle T\rangle^{\sigma_f\sigma_i}= \Tilde{Q}_0^{\sigma_f\sigma_i}(s\rightarrow 0)$ and $\langle T^2\rangle^{\sigma_f\sigma_i}=2\frac{\partial \Tilde{Q}_0^{\sigma_f\sigma_i}(s)}{\partial s}\Big|_{s\rightarrow 0}$ (whose expressions are lengthy, and dealt using {\bf Mathematica}). These are used to check the general condition given by Eq.~(\ref{trn_condition}).


\section{Simulation method to obtain ORR vanishing transition directly, or through evaluation of joint survival probabilities}\label{A4}

To study problems which have states with non-linear potentials, exact solutions are difficult to find in the presence of toggling and resetting, and the following numerical method was used in our work.  

We perform a hybrid simulation which is partly kinetic Monte Carlo (Gillespie algorithm \cite{Gillespie_1977}) and partly Langevin.  For the Gillespie part, if the motion starts in $+$ state, either the resetting event or a toggling is chosen with joint rate $(r + k_{off})$, while from  the $-$ state, the joint rate of resetting or toggling is $(r + k_{on})$. Consider the diffusing particle subjected to either potential $V_+(x)$ or $V_-(x)$ in $+$ or $-$ states respectively. So in between discrete Gillespie events, we evolve the particle using the continuous Langevin update rule
\begin{eqnarray}
    x_{n+1}= x_n -V_\pm'(x)dt+\sqrt{2Ddt}\ \xi
\end{eqnarray}
depending on the current state $\pm$. We choose $dt = 10^{-4}$, and $\xi$ is a Gaussian white noise with zero mean and delta correlation. Note the initial and reset position is $x_0 = 1$ and first passage happens when the particle reaches $x=0$. 



We used two methods to obtain ORR vanishing transition boundaries in parameter space which are discussed below.


\subsection{Method I: Obtaining the conditional probabilities in the absence of resetting, and associated moments, and using in the mathematical condition of ORR transition}

After updating the system using hybrid simulation method discussed above but without resetting ($r=0$), at any time $t$ we keep track of the number of trajectories $N(\sigma_f,t)$ which have survived up to time $t$ (i.e. $x > 0$) and in which the particle is in state $\sigma_f$, out of the total $N_{tot}$ trajectories studied in the simulation. Note the initial state in the beginning of the simulation is $\sigma_i$. 
Then 
\begin{eqnarray}
    Q_0^{\sigma_f\sigma_i}(t)=\frac{N(\sigma_f,t)}{N_{tot}}.
\end{eqnarray}

Using the above evaluated conditional probabilities, we obtained the desired conditioned moments of the times as follows by numerical integration: 
\begin{eqnarray}
    \langle T\rangle^{\sigma_f\sigma_i}=\int_0^\infty Q_0^{\sigma_f\sigma_i}(t)dt,
\end{eqnarray}
and
\begin{eqnarray}
    \langle T^2\rangle^{\sigma_f\sigma_i}=2\int_0^\infty tQ_0^{\sigma_f\sigma_i}(t)dt.
\end{eqnarray}
After calculating all the above moments we used those in the mathematical condition Eq. (\ref{trn_condition}) in the main text,
and located the ORR vanishing transition boundaries in parameter space. In particular, this method gave the filled circles in Fig. \ref{f6}, in the main text. 

\subsection{Method II: Obtaining MFPT as a function of $r$, and ORR vanishing transition thereafter}

After updating the system using hybrid simulation method discussed above, now with resetting ($r \neq 0$), we found first passage events and estimated MFPTs as a function of $r$. From such plots, we got the values of ORR, and hence obtained the parameter values  for which ORR vanishes.   In particular, this method gave the empty circles in Fig. \ref{f6}, in the main text.
\bibliography{reference}
\end{document}